\documentclass[onecolumn,amsmath,pra,superscriptaddress,nobibnotes,showpacs]{revtex4}
\usepackage{graphicx}
\begin{document}
\title{Effects of mismatched transmissions on two-mode squeezing and EPR correlations with a slow light medium}

\author{Sulakshana Thanvanthri}
\email{su1@umbc.edu} \affiliation{Department of Physics, University
of Maryland, Baltimore County, Baltimore, Maryland, 21250}
\author{Jianming Wen}
\affiliation{Department of Physics, University of Maryland,
Baltimore County, Baltimore, Maryland, 21250}
\author{Morton H. Rubin}
\affiliation{Department of Physics, University of Maryland, Baltimore County, Baltimore, Maryland, 21250}
\date{\today}

\begin{abstract}
We theoretically discuss the preservation of squeezing and
continuous variable entanglement of two mode squeezed light when the
two modes are subjected to unequal transmission. One of the modes is
transmitted through a slow light medium while the other is sent
through an optical fiber of unit transmission. Balanced homodyne
detection is used to check the presence of squeezing. It is found
that loss of squeezing occurs when the mismatch in the transmission
of the two modes is greater than 40$\%$ while near ideal squeezing
is preserved when the transmissions are equal. We also discuss the
effect of this loss on continuous variable entanglement using strong
and weak EPR criteria and possible applications for this
experimental scheme.

\end{abstract}


\maketitle

\section{Introduction}
The phenomenon of Electromagnetically Induced Transparency(EIT) has
been the focus of intensive research in the last decade. The
experiments performed so far using EIT have led to an extensive
understanding of the transmission of light pulses and preservation
of classical coherence features such as pulse shape. With the
emergence of the field of quantum computing, the ability to use EIT
to hold and transmit quantum coherence (and hence quantum
information) is a possible tool for creating future quantum
computational devices.

Squeezing of interferometric noise below the standard quantum limit
is an important quantum coherence phenomenon. Further two-mode
squeezing has been widely used to study continuous variable
entanglement. A recent experiment~\cite{akamatsu} examines the
preservation of squeezing when squeezed vacuum light is transmitted
through an EIT medium. In that experiment both modes of a two-mode
squeezed light were transmitted through an EIT medium and two mode
squeezing was maintained in the transmitted light over the entire
transparency window bandwidth. Another recent work ~\cite{peng}
theoretically, examines the preservation of single mode squeezing
and entanglement from an EIT system. In this paper we theoretically
examine the preservation of continuous variable entanglement and
squeezing when one of the modes of a two-mode squeezed light is
transmitted through an EIT medium while the other mode is propagated
through an optical fiber delay line. The light transmitted through
the EIT medium and the delay line is then be mixed together and
balanced homodyne measurements made to check for squeezing. This
experiment leads to a better understanding of the effects of the EIT
medium on the fluctuations, and hence coherence properties, of
quantum light. We find that differences in transmission of the two
modes leads to loss of entanglement and squeezing. We discuss the
possible causes for this loss and the effect of mismatched
transmission on the entanglement and EPR correlations
~\cite{einstein} in an ideal two-mode squeezed state. Possible
applications of this experimental scheme to continuous variable
teleportation and cryptographic schemes are also discussed.

\section{Experimental setup}

\begin{figure}[h]
\label{set-up}
\includegraphics [width=5.0in]{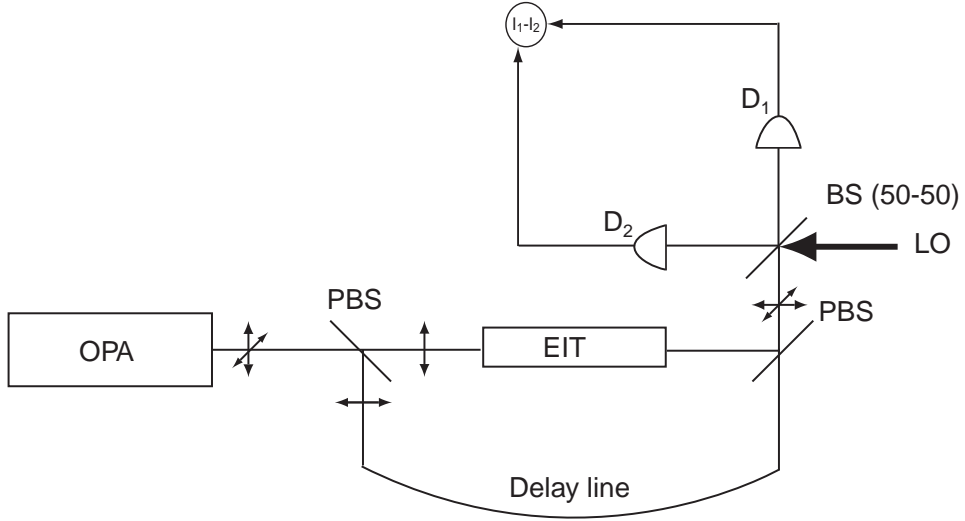}
\caption{\protect Experimental set-up to check preservation of squeezing in EIT medium }
\end{figure}

Figure 1 is the schematic experimental set-up. An Optical Parametric
Amplifier(OPA) pumped by a laser is the source of squeezed light.
The OPA uses a Type-II degenerate downconversion to generate
squeezed vacuum light in two orthogonally polarized modes with the
same frequency. The center frequency of the OPA is matched with the
center frequency of the probe transition of the EIT medium. The two
modes from the OPA are separated using a polarizing beam splitter
(PBS) and one of the modes is input into the EIT cell while the
other is sent through a delay line. The output from the EIT and the
delay line are mixed together at another polarizing beam splitter.
This light is then mixed with a local oscillator of the same
frequency at a 50-50 beam splitter(BS) and reaches the detectors
$D_{1}$ and $D_{2}$. The power spectrum of the difference current
from the two detectors gives the measure of the noise in the
quadrature measured. The control field of the EIT is external to the
setup and is assumed to be stable.

\section{Theoretical analysis}

In this section we provide a theoretical analysis of the experiment. We give a brief description of the OPA and
the EIT medium before entering into the complete calculation.

\subsection{Input-Output equations for OPA}

 The treatment of the OPA follows that in ~\cite{Ou}. We assume a cavity OPA with no input in the squeezed
 modes and an undepleted pump. The operator relations of the input and output fields of the OPA are as follows.
 The quantized electric field at the output of the OPA is given by
\begin{equation}\label{E-out-OPA}
    \mathbf{E}^{(+)}_{out}(t) = \sqrt{\frac{\hbar \omega_{0}}{4 \pi \epsilon_{0}c A}}
    (d^{out}_{o}(t)\mathbf{i}+d^{out}_{e}(t)\mathbf{j})
\end{equation}
where $\mathbf{i}$ and $\mathbf{j}$ are unit orthogonal vectors
indicating the polarization of the field. Note that we use continuum
field quantization for fields outside the OPA ~\cite{loudon}. $A$ is
the cross-sectional area determined by the geometry of the
experiment, $\omega_{0}$ is the center frequency of the OPA,
$d^{out}_{o}(t)$ and $d^{out}_{e}(t)$ are the annihilation operators
for the ordinary and extraordinary polarized photons at the output
of the OPA in a frame rotating with frequency $\omega_{0}$:
\begin{equation}\label{ft-dout}
d^{out}_{o, e}(t) =  \frac{1}{\sqrt{2\pi}} \int d\omega
D^{out}_{o,e}(\omega)e^{-i (\omega_{0}-\omega) t}.
\end{equation}
The integral in Eq.(\ref{ft-dout}) is over the bandwidth of the OPA.
 The relation between the input and
output operators of the OPA are given by
\begin{eqnarray}\label{in-out-eqns}
 D^{out}_{o}(\omega) & = & G(\omega) D^{in}_{o}(\omega)+g(\omega) D^{in \dag}_{e}(-\omega)\\ \nonumber
D^{out \dag }_{e}(-\omega) & = & G(\omega) D^{in \dag
}_{e}(-\omega)+ g(\omega)D^{in}_{o}(\omega)
\end{eqnarray}
where
\begin{eqnarray}\label{G-g-M}
    G(\omega) & = & [\kappa^{2}+(\gamma/2+i\omega)(\gamma/2-i\omega)]/M \\ \nonumber
    & = & G^{\ast}(-\omega) \\ \nonumber
    g(\omega) & = & \kappa\gamma/M  \\ \nonumber
    & =& g^{\ast}(-\omega)\\ \nonumber
    M & = & (\gamma/2-i\omega)^{2}-\kappa^{2}
\end{eqnarray}

\noindent $\gamma$ is the damping rate of the OPA, assumed to be the
same for both the modes, and $\kappa$ is the coupling coefficient
including the pump field of the OPA. For an amplifier $\kappa <
\gamma/2$. The bandwidth of the OPA is assumed to be $\gamma$. Note
that in this description of the OPA we have assumed that there are
no other external losses except through one port which serves as
both input and output port of the OPA.

\subsection{Parameters of EIT medium}
\begin{figure}[h]
\label{energy-scheme}
\includegraphics [width=3.0in]{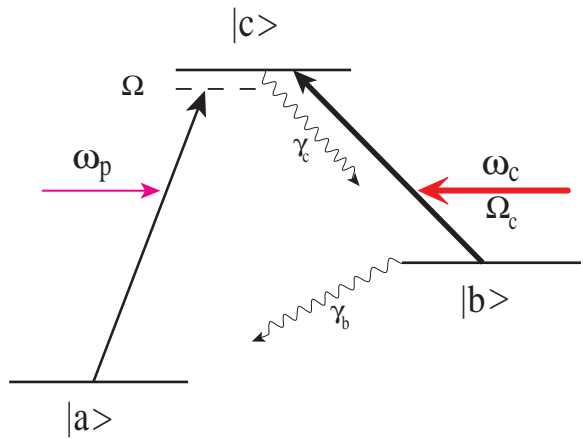}
\caption{\protect Energy level scheme for probe and control
transitions in an EIT medium}
\end{figure}

The energy level scheme for the one photon resonant EIT used in this
calculation is shown in Fig(2). In the one photon resonant EIT the
probe field ($\omega_{p}$) is detuned from the resonant frequency
($\omega_{ca}$) by $\Omega=\omega_{ca}-\omega_{p}$ and the control
field is assumed resonant. The susceptibility of the medium, as a
function of the detuning can be written as~\cite{scully},
\begin{equation}\label{chi-omega}
  \chi(\Omega) =\frac{N |\mu_{ac}|^{2}}{\hbar\epsilon_{0}}
  \frac{\Omega-i\gamma_{b}}{(\Omega-i\gamma_{b})(\Omega-i\gamma_{c})-\Omega_{c}^{2}}
\end{equation}
where $N$ is the density of the atoms in the EIT medium,
$\mu_{ac}$ is the dipole matrix element of the probe transition,
$\Omega_{c}$ is the Rabi frequency of the control field,
$\gamma_{b}$ and $\gamma_{c}$ are the decay rates of the states
$|b\rangle$ and $|c\rangle$ respectively.

 The complex factor picked up by the probe field as it traverses the medium is given by
\begin{eqnarray}\label{trans-coeff}
  T(\omega_{p})=& e^{ik_{EIT}(\omega_{p})z}=&  e^{i(k_{EIT}(\omega_{ca})-\Omega\frac{dk_{EIT}}{d\omega}|_{\omega_{ca}}+
  \Omega^{2}\frac{d^{2}k_{EIT}}{d\omega^{2}}|_{\omega_{ca}})z}
\end{eqnarray}
where $z$ is the length of the EIT medium. We have made an
assumption that the detuning from resonance $\Omega \ll \omega_{ca}$
so that the wave vector $k_{EIT}$ can be expanded about the
resonance. Using
\begin{equation}\label{k-eit}
k_{EIT}(\omega)= \frac{\omega}{c}\sqrt{1+\chi(\omega)}
\end{equation}
and assuming that $|\chi(\omega)|\ll 1$, we get,
\begin{equation}\label{T-Omega}
    T(\Omega)=e^{i\frac{\omega_{ca}z}{c}} e^{\frac{-\chi_{2}(\Omega)\omega_{ca}z}{2c}}e^{i\frac{\chi_{1}(\Omega)\omega_{ca}z}{2c}}
\end{equation}
 where $\chi_{1}$ and $\chi_{2}$ are the real and imaginary parts
of the susceptibility.

\subsection{Calculation}

We now proceed to calculate the noise in the measured quadrature of
light transmitted through the EIT and the delay line. We use a
circularly polarized classical local oscillator field,
\begin{equation}\label{lo-field}
  \mathbf{E}_{LO}(t)=
  |\beta_{LO}|(e^{-i(\Omega_{LO}t-\phi_{LO})}+e^{i(\Omega_{LO}t-\phi_{LO})})(\mathbf{i}+\mathbf{j})
\end{equation}
where $|\beta_{LO}|$ is the classical amplitude, $\Omega_{LO}$ the
frequency and $\phi_{LO}$ the tunable phase of the local
oscillator field. In the following calculation, we assume that the
central frequency of the OPA ($\omega_{o}$), the central frequency
of the probe transition ($\omega_{ca}$) and the local oscillator
frequency ($\Omega_{LO}$) are identical.
\begin{equation}\label{freq-equal}
\omega_{o}\equiv\omega_{ca}\equiv\Omega_{LO}
\end{equation}

The field at the detectors $D_{1}$ and $D_{2}$ are given by~\cite{glauber},
\begin{eqnarray}\label{detect-fields}\
\mathbf{E}_{1}^{(+)}(t) & = & \frac{i}{\sqrt{2}} \sqrt{\frac{\hbar
\omega_{0}}{4 \pi \epsilon_{0}c A}} (i
d_{A}(t) \mathbf{i}+d_{B}(t) \mathbf{j})+\frac{1}{\sqrt{2}}|\beta_{LO}|e^{-i(\Omega_{LO}t-\phi_{LO})}(\mathbf{i}+ \mathbf{j})\\
\mathbf{E}_{2}^{(+)}(t) & = & \frac{1}{\sqrt{2}} \sqrt{\frac{\hbar
\omega_{0}}{4 \pi \epsilon_{0}c A}} (i
d_{A}(t)\mathbf{i}+d_{B}(t)\mathbf{j})+\frac{i}{\sqrt{2}}|\beta_{LO}|e^{-i(\Omega_{LO}t-\phi_{LO})}(\mathbf{i}+
\mathbf{j})
\end{eqnarray}
$d_{A}(t)$ and $d_{B}(t)$ are photon annihilation operators at the
output of the EIT and the delay line respectively. Homodyne
detection involves studying the noise in the difference current from
the two detectors~\cite{scully}. The difference current is given by
\begin{eqnarray}\label{i-diff}
I_{D}(t) & = &
\mathbf{E}_{2}^{(-)}(t)\cdot\mathbf{E}_{2}^{(+)}(t)-\mathbf{E}_{1}^{(-)}(t)\cdot\mathbf{E}_{1}^{(+)}(t) \\
\label{i-diff-1 } & = & i  \sqrt{\frac{\hbar \omega_{0}}{4 \pi
\epsilon_{0}c A}}|\beta_{LO}|
\left[e^{-i(\Omega_{LO}t-\phi_{LO})}(-i
d_{A}^{\dag}(t)+d_{B}^{\dag}(t))-e^{i(\Omega_{LO}t-\phi_{LO})}(i
d_{A}(t)+d_{B}(t))\right]
\end{eqnarray}

Since $\Omega_{LO}=\omega_{o}$, we use the transformation,
\begin{equation}\label{d_to_D}
d_{A}(t)e^{i\Omega_{LO}t} = D_{A}(t)
\end{equation}
$I_{D}(t)$ can be re-written as
\begin{equation}\label{Id_D}
 I_{D}(t)=i  \sqrt{\frac{\hbar \omega_{0}}{4 \pi \epsilon_{0}c A}}|\beta_{LO}| \left[e^{i\phi_{LO}}(-i
D_{A}^{\dag}(t)+D_{B}^{\dag}(t))-e^{-i\phi_{LO}}(i
D_{A}(t)+D_{B}(t))\right]
\end{equation}

We now Fourier transform Eq.(\ref{Id_D}) since we are interested in
the spectrum of squeezing.

\begin{equation} \label{ft-ID}
I_{D}(\Omega)=i  \sqrt{\frac{\hbar \omega_{0}}{4 \pi \epsilon_{0}c
A}}|\beta_{LO}| \left[e^{i\phi_{LO}}(-i
D_{A}^{\dag}(-\Omega)+D_{B}^{\dag}(-\Omega))-e^{-i\phi_{LO}}(i
D_{A}(\Omega)+D_{B}(\Omega))\right]
\end{equation}

The operators $D_{A}(\Omega)$ and $D_{B}(\Omega)$ are now expressed
in terms of operators at the output of the OPA.
\begin{eqnarray}\label{DA_to_Dout}
D_{A}(\Omega) & = & T(\Omega)D_{o}^{out}(\Omega)+ i R(\Omega) D_{v}(\Omega)\\
D_{B}(\Omega) & = & i e^{ik_{f}(\Omega)l_{f}}D_{e}^{out}(\Omega)=
ie^{i(\frac{n_{f}(\omega_{0})\omega_{0}}{c}-\frac{\Omega}{V_{f}})l_{f}}D_{e}^{out}(\Omega)=ie^{i\phi_{d}(\Omega)}D_{e}^{out}(\Omega)
\end{eqnarray}
We have used a beam splitter to model the loss due to the EIT
medium~\cite{scully}. Note that the amplitude transmittance and
reflectance, $T(\Omega)$ and $R(\Omega)$, are complex quantities. $
D_{v}(\Omega)$ is the annihilation operator of an uncorrelated
vacuum mode entering into the system,
 $n_{f}(\omega_{0})$ is the refractive index of the fiber at the
central frequency of the light, $V_{f}$ is the group velocity of the
light in the fiber and $l_{f}$ is the length of the fiber. We assume
a lossless fiber and ignore the dispersion in the fiber. The light
through the EIT is filtered by the transparency window and the light
transmitted through the delay line picks up a phase
$\phi_{d}(\Omega)$ where $l_{f}$ is chosen such that the fields at
the output of the EIT and the fiber arrive in phase.
Eq.(\ref{ft-ID}) is rewritten as
\begin{eqnarray}\label{ID_Dout}
  I_{D}(\Omega)& = & \sqrt{\frac{\hbar \omega_{0}}{4 \pi \epsilon_{0}c A}}|\beta_{LO}| \left[e^{i\phi_{LO}}(T^{\ast}(-\Omega) D_{o}^{out
\dag}(-\Omega)-i R^{\ast}(-\Omega) D_{v}^{\dag}(-\Omega)\right.\\
\nonumber & & \left.+e^{-i\phi_{d}(-\Omega)}D_{e}^{out
\dag}(-\Omega))+e^{-i\phi_{LO}}( T(\Omega)D_{o}^{out}(\Omega)+i
R(\Omega)
D_{v}(\Omega)+e^{i\phi_{d}(\Omega)}D_{e}^{out}(\Omega))\right]\\
\label{quadrature}
 & = & 2\sqrt{\frac{\hbar \omega_{0}}{ \pi \epsilon_{0}c A}}|\beta_{LO}|X^{\phi_{LO}}(\Omega)
\end{eqnarray}
The quadrature being measured, $X^{\phi_{LO}}(\Omega)$,  is defined
in Eq.(\ref{quadrature}).

We can use Eqs.(\ref{in-out-eqns}) to relate the output operators to
the vacuum input of the OPA. The presence of squeezing is checked by
measuring the noise level of the quadrature and comparing with
standard quantum level, in this case, 0.5. The result of this
calculation is found to be
\begin{eqnarray}\label{del_x}
  (\Delta X^{\phi_{LO}}(\Omega))^{2}& = &
  \frac{1}{4}\left[|R(\Omega)|^{2}+|G(\Omega)|^{2}(1+|T(\Omega)|^{2}) +
  |g(\Omega)|^{2}(1+|T(-\Omega)|^{2})\right. \\ \nonumber
  & & \left. +2|T(\Omega)| Re(g(\Omega)G^{\ast}(\Omega)
  e^{i(\phi_{d}(-\Omega)-2\phi_{LO}+\frac{\chi_{1}(\Omega)\omega_{0}z}{2c}+\frac{\omega_{0}z}{c})})\right. \\
  \nonumber
  & & \left.+ 2|T(-\Omega)| Re(g(\Omega)G^{\ast}(\Omega)
  e^{i(\phi_{d}(\Omega)-2\phi_{LO}+\frac{\chi_{1}(-\Omega)\omega_{0}z}{2c}+\frac{\omega_{0}z}{c})})\right]
\end{eqnarray}
We make the approximations $|T(-\Omega)|=|T(\Omega)|$ since
$\chi_{2}(-\Omega)=\chi_{2}(\Omega)$) and
$\chi_{1}(-\Omega)=-\chi_{1}(\Omega)$ \cite{scully}. These
approximations are valid only within the region of transparency and
when the control field is resonant to the atomic transition
$|b\rangle\rightarrow|c\rangle$. Using
$|R(\Omega)|^{2}+|T(\Omega)|^{2}=1$, we have
\begin{eqnarray}\label{del_x_approx}
  (\Delta X^{\phi_{LO}}(\Omega))^{2}& = &
  \frac{1}{4} [(1-|T(\Omega)|^{2})+(|G(\Omega)|^{2}+|g(\Omega)|^{2})(1+|T(\Omega)|^{2})+
  4|T(\Omega)|g(\Omega)G^{\ast}(\Omega)\times \\ \nonumber
  & & cos(\frac{\Omega}{V_{f}}l_{f}+\frac{\chi_{1}(\Omega)\omega_{0}z}{2c})
   cos(2\phi_{LO}-\frac{\omega_{0}z}{c}-\frac{n_{f}(\omega_{0})\omega_{0}}{c}l_{f})]
\end{eqnarray}

This is the final expression for the noise in the phase quadrature
measured.The cosine terms in the above equation determine the phase
quadrature being measured. The second cosine term varies only with
the local oscillator phase $\phi_{LO}$ since
$\frac{\omega_{0}z}{c}+\frac{n_{f}(\omega_{0})\omega_{0}}{c}l_{f}$,
the first order phase picked up due to the EIT and the delay line,
is a constant for the experiment. The first cosine term depends on
phase terms that are first order in the detuning $\Omega$, for the
values of EIT medium parameters used in this calculation, and the
fiber. $\frac{\chi_{1}(\Omega)\omega_{0}z}{2c}$ is the delay due to
the EIT medium and,  $\frac{\Omega}{V_{f}}l_{f}$ is the compensating
delay introduced by the fiber. Note that though the phase quadrature
$X^{\phi_{LO}}$ is also determined by the Rabi frequency of the
control field through $\chi_{1}(\Omega)$, the squeezed quadrature
remains unaffected as long as the fluctuations in the control field
amplitude are not large.

 The following were the values of OPA parameters used in the
calculation whose results are illustrated in Fig.3:

The damping rate of the OPA is modeled as the leakage due to the
input-output coupling mirror and given by $\gamma = ct/2L$ where
$t$ is the transmittance of the mirror, and $L$ is the length of
the OPA cavity. $\gamma$ is also a rough estimate of the bandwidth
of squeezed light from the OPA. The values of the above parameters
are chosen such that the bandwidth of the OPA is about 40 MHz,
much larger then the transparency window. $\kappa$, the coupling
coefficient is chosen to be below the threshold of an oscillator,
$\kappa = (0.6)\gamma/2$

The EIT medium is Rubidium(Rb) vapor and the probe resonance is
the $D_{1}$ line of Rb at 795 nm. The other EIT and optical fiber
parameters have the following values:

$|\Omega_{c}| = 20$MHz, $N=2.7 \times 10^{17}$ atoms/m$^{3}$,
$z=0.05$m, $\gamma_{b}=10^{4}$Hz, $\gamma_{c}=6\pi\times10^{6}$Hz
and $|\mu_{ac}|=1.46 \times 10^{-29}$. We assumed $n_{f}=1.5$ for
frequencies around the probe resonance and $V_{f}=10^{8}m/s$ for the
group velocity of light in the fiber. This gives us the length of
the fiber $l_{f}=3.04$ km. The transparency window is approximately
1.2 MHz for the given control field strength.

\section{Results and Discussion}

In order to clearly understand the effects of mismatched
transmissions on quantum coherence it is useful to compare with the
case of equal transmissions of both modes of light.

By inspection of Eq.(\ref{del_x_approx}), the expression for the
quadrature variance for identical transmission of the two modes is
given by,

\begin{eqnarray}\label{del_x_approx-2}
  (\Delta X^{\phi_{LO}}(\Omega))^{2}& = &
  \frac{1}{2} [(1-|T(\Omega)|^{2})+(|G(\Omega)|^{2}+|g(\Omega)|^{2})|T(\Omega)|^{2}+
  2|T(\Omega)|^{2}g(\Omega)G^{\ast}(\Omega)\times \\ \nonumber
  & & cos(\frac{\Omega}{V_{f}}l_{f}+\frac{\chi_{1}(\Omega)\omega_{0}z}{2c})
   cos(2\phi_{LO}-\frac{\omega_{0}z}{c}-\frac{n_{f}(\omega_{0})\omega_{0}}{c}l_{f})].
\end{eqnarray}

\begin{figure}[h]
\label{figures}
\includegraphics [height=2.29in]{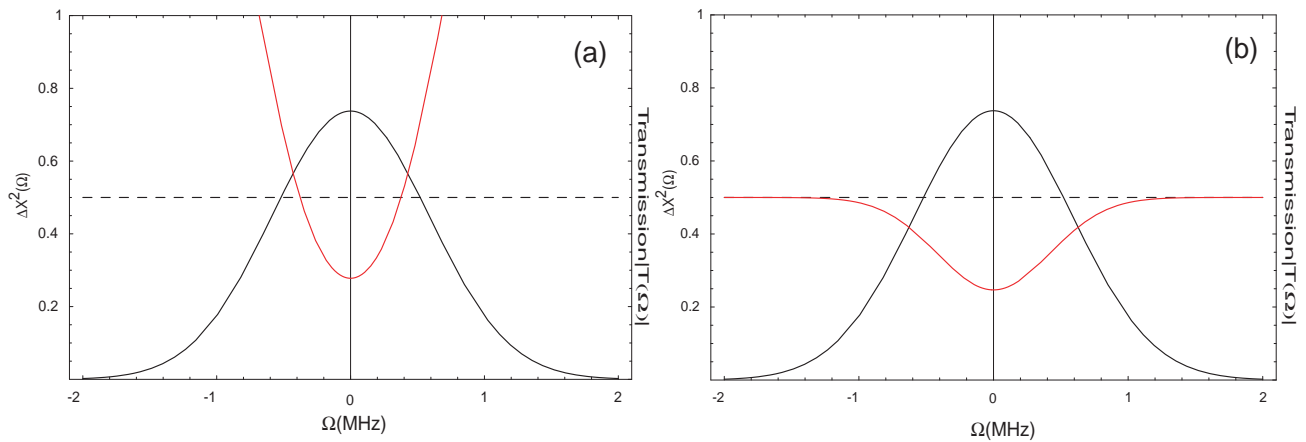}
\caption{$(\Delta X)^{2}$(red) and $|T(\Omega)|$(black) as a
function of detuning $\Omega$ ($\phi_{LO}$=2.4). The dashed line
$|T(\Omega)|=0.5$ defines the bandwidth of the transparency window
and the standard quantum noise limit $(\Delta X)^{2}=0.5$. Squeezing
is indicated when the quadrature noise falls below the dashed line.
In (a), the transmissions of the modes are not equal and the
squeezing bandwidth is less than the transparency bandwidth. In (b),
the transmissions are equal and near ideal squeezing is preserved.}
\end{figure}

In Fig.3(a) and (b), we plot the quadrature variance derived in
Eqs.(\ref{del_x_approx}) and (\ref{del_x_approx-2})  and
$|T(\Omega)|$ as a function of detuning. When the transmissions are
mismatched squeezing is maintained over a bandwidth less than the
transparency window. For matched transmissions, near ideal squeezing
is maintained even beyond the transparency bandwidth defined by the
$|T(\Omega)|= 0.5$ line. In ~\cite{akamatsu},the two modes in that
experiment have different frequencies. Squeezing is preserved over
the entire transparency bandwidth but not beyond. We believe that
this is because of nearly equal complex transmissions of the two
modes. The difference in phase, picked up in the EIT medium by the
two modes with slightly different frequencies, spoils the two-mode
squeezing beyond the transparency bandwidth.

Quadrature phase amplitude measurements of two mode squeezed light
have often been used to study continuous variable entanglement
~\cite{Ou, Reid1}. The experiment discussed in this paper can be
used  to study the effect of EIT transmission on two mode squeezing
and entanglement simultaneously although for a rigorous test of
entanglement, the two modes may be detected separately and
correlation measurements made on the photocurrents. Before  we
discuss these criteria, we need to rewrite the measured quadrature
as follows.

\begin{eqnarray}\label{quad_new}
   X^{\theta}_{\phi}(\Omega)&=
  &\frac{1}{2}\left[(e^{i\theta}a^{\dagger}(-\Omega)+e^{-i\theta}a(\Omega))
  -(e^{i\phi}b^{\dagger}(-\Omega)+e^{-i\phi}b(\Omega))\right]\\
  & = & X_{a}-X_{b}.
\end{eqnarray}
where
\begin{eqnarray}\label{a-and-b}
a(\Omega) =T(\Omega)D_{o}^{out}(\Omega)+ i R(\Omega) D_{v}(\Omega),&
b(\Omega) = e^{i\frac{\Omega}{V_{f}}l_{f}}D_{e}^{out}(\Omega)
\\ \nonumber
\phi = \pi+\phi_{LO}-\frac{n_{f}(\omega_{0})\omega_{0}}{c}l_{f},
&\theta=\phi_{LO}.
\end{eqnarray}

$X^{\theta+\pi/2}_{\phi-\pi/2}(\Omega)$, the quadrature involving
the non-commuting observables of $X_{a}$ and $X_{b}$ can be written
as
\begin{eqnarray}\label{p_quad}
    X^{\theta+\pi/2}_{\phi-\pi/2}(\Omega)& = & \frac{i}{2}\left[(e^{i\theta}a^{\dagger}(-\Omega)-e^{-i\theta}a(\Omega))+
    (e^{i\phi}b^{\dagger}(-\Omega)-e^{-i\phi}b(\Omega))\right]\\
    & = &  P_{a}+P_{b}
\end{eqnarray}

Reid has suggested two EPR criteria - strong and weak - based on
phase amplitude measurements to detect entanglement and EPR
correlations in systems that are not maximally
entangled~\cite{Reid2}. These measurements are pertinent when an EIT
medium is used since even perfectly entangled states are bound to be
degraded by passage through the EIT medium as along as transmission
is not unity. The stronger criterion given by,
\begin{equation}\label{EPR}
    \Delta(X_{a}-X_{b})\Delta(P_{a}+P_{b})< \frac{1}{4},
\end{equation}
is in the spirit of the original EPR paradox~\cite{einstein} where
local realism is defined with no assumptions regarding the form of a
local realistic theory.

The weaker criterion, which is also a two-mode squeezing criterion,
is given by,
\begin{equation}\label{Entanglement}
    \Delta(X_{a}-X_{b})\Delta(P_{a}+P_{b})< \frac{1}{2}.
\end{equation}
This weaker criterion is based on the inseparability of an entangled
state. The entanglement criteria suggested by Simon~\cite{simon} and
Duan ~\cite{duan} also belong to the weaker type.

\begin{figure}[h]
\label{figures}
\includegraphics [height=2.5in]{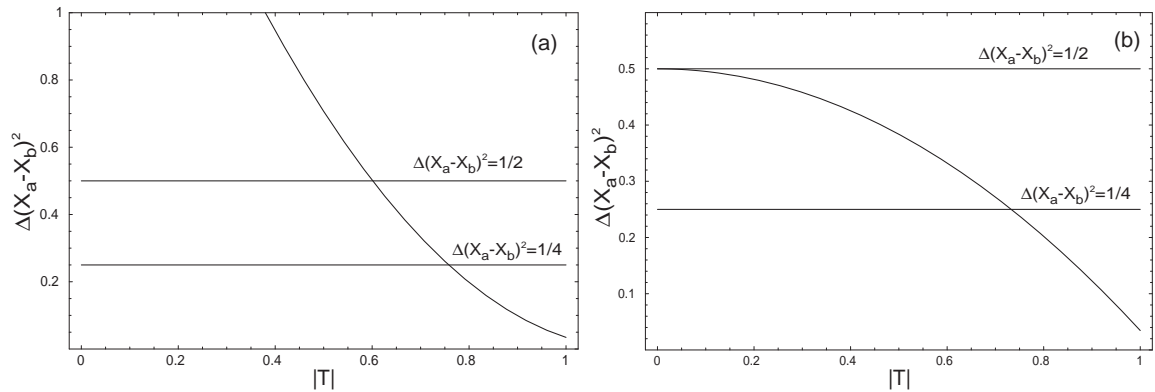}
\caption{$(\Delta X)^{2}$ as a function of transmission at
$\Omega=$50 KHz and $\phi_{LO}$=2.4. In (a) transmission of the two
modes are unequal and squeezing is maintained only for $|T|>0.5$. In
(b) both modes have equal transmission and all the transmitted light
is squeezed. In both cases as the transmission increases first the
weaker and then the stronger EPR criteria are satisfied.}
\end{figure}

Fig.4(a) shows the variance in one of the quadratures,
$\Delta(X_{a}-X_{b})^{2}$,  as a function of transmission through
the EIT medium for a particular detuning. From the expressions for
the phases $\theta$ and $\phi$ and the error in one of the
quadratures given by Eq.(\ref{del_x_approx}), we see that the
expression for quadrature noise does not change from
$X^{\theta}_{\phi}(\Omega)$ to
$X^{\theta+\pi/2}_{\phi-\pi/2}(\Omega)$. So both the quadratures,
$X_{a}-X_{b}$ and $P_{a}+P_{b}$, are squeezed for a particular
detuning $\Omega$ and obey the weaker entanglement criterion in
Eq.(\ref{Entanglement}) when the mismatch between the transmissions
is less than 40$\%$. With further increase in transmission, the
stronger criterion is also satisfied.

In Fig 4(b), where the transmissions are equal, all the transmitted
light is squeezed and entangled either weakly or strongly. The
squeezing is almost ideal. However, the strong EPR criteria is not
satisfied till transmission is about 0.75 though the light from the
OPA is ideally squeezed and obeys both the EPR criteria over the
entire bandwidth $\gamma$.

We find that the differences in transmission of the two modes
adversely affect both entanglement and squeezing. This can be
explained as follows. Squeezing occurs due to destructive
interference of phase sensitive and insensitive fluctuations. As
long as the transmissions of the two modes is not equal, the
magnitude of destructive interference is not enough for the light to
be squeezed. The low correlation of fluctuations also causes the
loss of entanglement.

Loss of light through absorption or filtering is a common occurrence
in any real world optical experiment. Up until now, the aim has been
to avoid such losses in case of squeezing experiments since they
lead to reduced squeezing efficiency by adding vacuum fluctuations.
In this calculation we see that for two-mode squeezed light,
squeezing can be preserved as long as the transmissions of the two
modes are matched, even if they are less than unity. By coupling one
mode into a fiber and varying it's transmission, we can maintain
two-mode squeezing in a small range of detuning within the
transparency window while destroying squeezing coherence for all
other detuning frequencies. Here we have used an EIT medium since it
provides a smooth well understood transmission feature over a
compact frequency bandwidth.

Teleportation and cryptography schemes using continuous variable
entanglement in squeezed states have long been discussed in
literature~\cite{Ralph, Reid3}. The ability to selectively preserve
entanglement and squeezing can be very useful in such schemes. The
experiment suggested here provides a practical method to do this.
For example, if the two modes are shared between two parties, the
sequence of fiber transmissions, required to reconstruct the
information contained in the entanglement, could be the shared key.
In the absence of the key the information cannot be reliably
reconstructed. Further any eavesdropping, by diverting part of a
transmitted mode, adds vacuum noise that is easily detectable if the
message cannot be reconstructed with the shared key. The EIT medium,
apart from providing a varying transmission also slows down the
transmitted light. The effects discussed here may also be important
for application of any slow light effects in such continuous
variable teleportation or cryptographic schemes.

\section{Summary}

In this paper we have theoretically investigated the preservation of
two mode squeezing and entanglement when the transmission of the two
modes are not equal. An EIT medium provides varying transmission for
one the modes and a fiber is used to transmit the other mode. We
find that two mode squeezing is preserved as the transmission of
both modes approach the same value. Using the strong and weak
continuous variable entanglement criteria, we find that entanglement
in the weak regime is maintained whenever two mode squeezing is
preserved but loss of transmission degrades entanglement in the
strong regime. We have also outlined a scheme by which two mode
squeezing and entanglement can be selectively maintained or
destroyed using varying fiber transmissions. Such a scheme, if
experimentally demonstrated, can have interesting application to
continuous variable quantum cryptography.

\section{Acknowledgements}

We wish to thank Yanhua Shih and members of the UMBC Quantum Optics
group for helpful discussions of the material presented in this
paper. We thank Daisuke Akamatsu, Physics Department, Tokyo
Institute of Technology, for details regarding the experimental
results of his group. This work was partially supported by Naval
Research Laboratory contract N00173-04-P-1311. We also wish to thank
the referee for pointing out an error in the earlier version of this
paper.

\end{document}